# SIMULATION OF FRACTURE COALESCENCE IN GRANITE VIA THE COMBINED FINITE-DISCRETE ELEMENT METHOD

*B. Euser[1,*], E. Rougier[1], Z. Lei[1], E. E. Knight[1], L. P. Frash[2], J. W. Carey[2], H. Viswanathan[3], A. Munjiza[4]*

*[1]Los Alamos National Laboratory - Geophysics Group*
*[2]Los Alamos National Laboratory – Earth Systems Observations Group*
*[3]Los Alamos National Laboratory - Computational Earth Science Group*
*[4]FGAG – University of Split, Croatia*

[*] Corresponding Author: Dr. Bryan Euser, beuser@lanl.gov

## ABSTRACT

Fracture coalescence is a critical phenomenon for creating large fractures from smaller flaws, affecting fracture network flow and seismic energy release potential. In this paper, simulations of fracture coalescence processes in granite specimens with pre-existing cracks are performed. These simulations utilize an in-house implementation of the Combined Finite-Discrete Element method (FDEM) known as the Hybrid Optimization Software Suite (HOSS). The pre-existing cracks within the specimens follow two geometric patterns: 1) a single crack oriented at different angles with respect to the loading direction, and 2) two cracks, where one crack is oriented perpendicular to the loading direction and the other crack is oriented at different angles. The intent of this study is to demonstrate the suitability of FDEM for modeling fracture coalescence processes including: crack initiation and propagation, tensile and shear fracture behavior, and patterns of fracture coalescence. The simulations provide insight into the evolution of fracture tensile and shear fracture behavior as a function of time. The single-crack simulations accurately reproduce experimentally measured peak stresses as a function of crack inclination angle. Both the single- and double-crack simulations exhibit a linear increase in strength with increasing crack angle; the double-crack specimens are systematically weaker than the single-crack specimens.

## 1. INTRODUCTION

Natural rock contains a large number of natural fractures that can range in scale from sub-micron to kilometers in length [1]. A thorough understanding of crack initiation, propagation, interaction, and eventual coalescence emanating from pre-existing fractures is key to characterizing geologic risk in engineering design and to improving the safety and effectiveness of subsurface hydrologic, energy, and waste disposal activities [2–4]. In an effort to better understand the fracture processes within brittle solids (i.e., rock-like materials), crack coalescence has been extensively studied, both experimentally and numerically [5–16].

Many numerical methods have been employed in the simulation of crack initiation and propagation, such as the finite element method (FEM) [17, 18], boundary element method (BEM) [19–22], and discrete element method (DEM) [23, 24]. However, these methods all present various deficiencies when trying to reproduce fracture and fragmentation processes in geologic materials. For example, FEM cannot precisely specify the location of a crack; the finest resolution of the model is that of the mesh (i.e., an element can be flagged as "cracked"), the location of the crack inside the element is unknown. Furthermore, FEM cannot accurately handle large deformations and fragmentation processes due to the continuum assumption (i.e., detached material is difficult to model) [25]. In contrast, DEM provides a natural framework to describe de-bonding (i.e., fracturing) among the discrete elements. However, in order to describe the bulk elastic and strength properties of a material a thorough calibration process (i.e., inverse tuning) of the modeling parameters needs to be conducted, making the connection between the model and the measured material properties more indirect [26]. In addition, DEM simulations require a lot of computational resources, limiting their applicability. The BEM has often been used in the study of fracture coalescence, though results are sometimes unsatisfactory [27]; additionally, incorporating the stress, strain, and energy into the failure criterion is not an inherent capability of BEM [27].

Within the context of rock mechanics, the Combined Finite-Discrete Element method (FDEM) has been applied to many complex industrial problems such as block caving, deep mining techniques (tunneling, pillar strength, etc.), rock blasting, seismic wave propagation, dam stability, rock slope stability, rock mass strength characterization problems, etc. [28–31]. The implementation of the combined finite-discrete element method [32–34] utilized in this study is known as HOSS (Hybrid Optimization Software Suite) has been specifically designed to bridge the gaps in FEM and DEM methods by directly coupling measurable parameters with separable elements in order to better interrogate fracture coalescence mechanisms. This paper presents a general overview of the FDEM methodology, a description of single- and double-crack laboratory scale experiments conducted by Lee and Jeon [23] and numerical models used to replicate experimental fracture coalescence processes.

Experimental studies provide a fundamental understanding of material failure mechanisms and crack coalescence modes while providing a basis for building conceptual frameworks to describe the nature of fracture initiation and propagation in brittle materials [13–15]. Understanding of fracture coalescence can be further enhanced by analysis developed through numerical modeling, which can also be used to bridge length scales (i.e., from laboratory-scale experiments to field scale problems). The present study is part of a broader effort to characterize fracture formation and permeability evolution in dynamically sheared rock. In earlier work, we conducted 3-D FDEM analysis of split-Hopkinson pressure bar experiment on granite in which the temporal evolution of strain and tensile stress was determined including a model of strain softening following initial specimen fracture. We have also applied 2-D FDEM analysis to the interpretation of triaxial direct shear permeability experiments on shale in which we have reproduced fracture patterns [2] and analyzed stress development and stress-strain history to specimen failure [3]. In these projects, HOSS is used to interpret mechanisms of fracture formation and fluid flow observed in experiments and to provide fracture patterns for use in larger-scale discrete fracture network models. The objective of this study is to demonstrate that the FDEM method as implemented in HOSS correctly assesses fracture coalescence processes.

## 2. PRINCIPLES OF FDEM

**FDEM Methodology.** FDEM combines a finite-element-based analysis of continua with discrete-element-based transient dynamics, contact detection, and contact interaction solutions. Solid domains (called discrete elements) are discretized into finite elements. Finite rotations and displacements are assumed *a priori* and are formulated using a multiplicative-decomposition-based finite strain formulation [34]. The finite element discretization of solid domains is conveniently used to discretize the contact between discrete elements. Utilizing this approach, discretized contact solutions are then used for resolving both contact detection and contact interaction [33, 35]. The general governing equation solved in FDEM is [33]:

$$\mathbf{M}\ddot{\mathbf{x}} + \mathbf{C}\dot{\mathbf{x}} = \mathbf{f} \tag{1}$$

where $\mathbf{M}$ and $\mathbf{C}$ are the lumped mass matrix and the damping matrix, respectively; $\mathbf{x}$ is the displacement vector, and $\mathbf{f}$ is the equivalent force vector acting on each node. Equation (1) is integrated with respect to time in order to obtain the transient evolution of the system.

Fracture and fragmentation processes are defined via the combined single and smeared discrete crack approach with the capability of directly incorporating experimentally determined stress-strain and stress-displacement data [36]. This approach decomposes the relationship between stress and strain into two regimes, which represent strain-hardening and strain-softening. In the strain-hardening regime, no discrete damage (i.e., fracture at the boundaries of the elements) occurs in the material and a continuum constitutive law is employed together with non-softening material nonlinearity (i.e., plasticity, etc.). The strain-softening portion of the constitutive law is represented via a description of the evolution of stresses at the interfaces as a function of relative displacement (not strain); with this approach, the "strain localization" portion of the constitutive law is introduced via a set of material parameters that describe the width of the damage zone associated with the strain localization band and/or discrete cracks.

**Damage.** Figure 1a illustrates how the cohesive stresses at the boundaries of the finite elements evolve with relative displacement. The parameters describing the stresses at the interface of the finite elements are the elastic threshold relative displacement $\delta^{\mathrm{e}}$, the maximum relative displacement $\delta^{\mathrm{e}} + \delta^{\max}$ and the maximum stress $\sigma^{\max}$. The strain hardening portion of the curve is defined by $0 \le \delta \le \delta^{e}$, while the strain softening portion of the curve is defined by $\delta^{\mathrm{e}} < \delta \le \delta^{\mathrm{e}} + \delta^{\max}$. The curve shown in Figure 1a is used to describe normal (i.e., pure tension) behavior by setting

$$\delta^{\mathrm{e}} = \delta^{\mathrm{e}}_{\mathrm{n}} \;\; ; \;\; \delta^{\max} = \delta^{\max}_{\mathrm{n}} \;\; ; \;\; \sigma^{\max} = \sigma^{\max}_{\mathrm{n}} \tag{2}$$

where $\delta^{\mathrm{e}}_{\mathrm{n}}$, $\delta^{\max}_{\mathrm{n}}$ and $\sigma^{\max}_{\mathrm{n}}$ are the elastic threshold, the maximum relative displacement, and the maximum stress in the normal direction respectively; or the tangential (i.e., pure shear) behaviour by setting

$$\delta^{\mathrm{e}} = \delta^{\mathrm{e}}_{\mathrm{t}} \;\; ; \;\; \delta^{\max} = \delta^{\max}_{\mathrm{t}} \;\; ; \;\; \sigma^{\max} = \sigma^{\max}_{\mathrm{t}} \tag{3}$$

where $\delta^{\mathrm{e}}_{\mathrm{t}}$, $\delta^{\max}_{\mathrm{t}}$ and $\sigma^{\max}_{\mathrm{t}}$ are the elastic threshold, the maximum relative displacement and the maximum stress in the tangential or in-plane direction, respectively (Figure 1b). Tensile fracture occurs when the elastic threshold displacement (and the corresponding maximum normal stress) is in the cohesive element is reached. Tangential fracture initiation occurs when the maximum tangential stress of the cohesive element is reached, where the tangential stress is defined as

$$\sigma_{\mathrm{t}} = c + \sigma_{\mathrm{n}} \tan \phi_{\mathrm{c}} \tag{4}$$

where $c$ is internal cohesion, $\sigma_n$ is the normal stress acting on the cohesive element, and $\phi_c$ is the internal angle of friction. Once the material exits the strain-hardening regime (i.e., reaches maximum shear/tensile stress) fracture and fragmentation processes begin to occur.

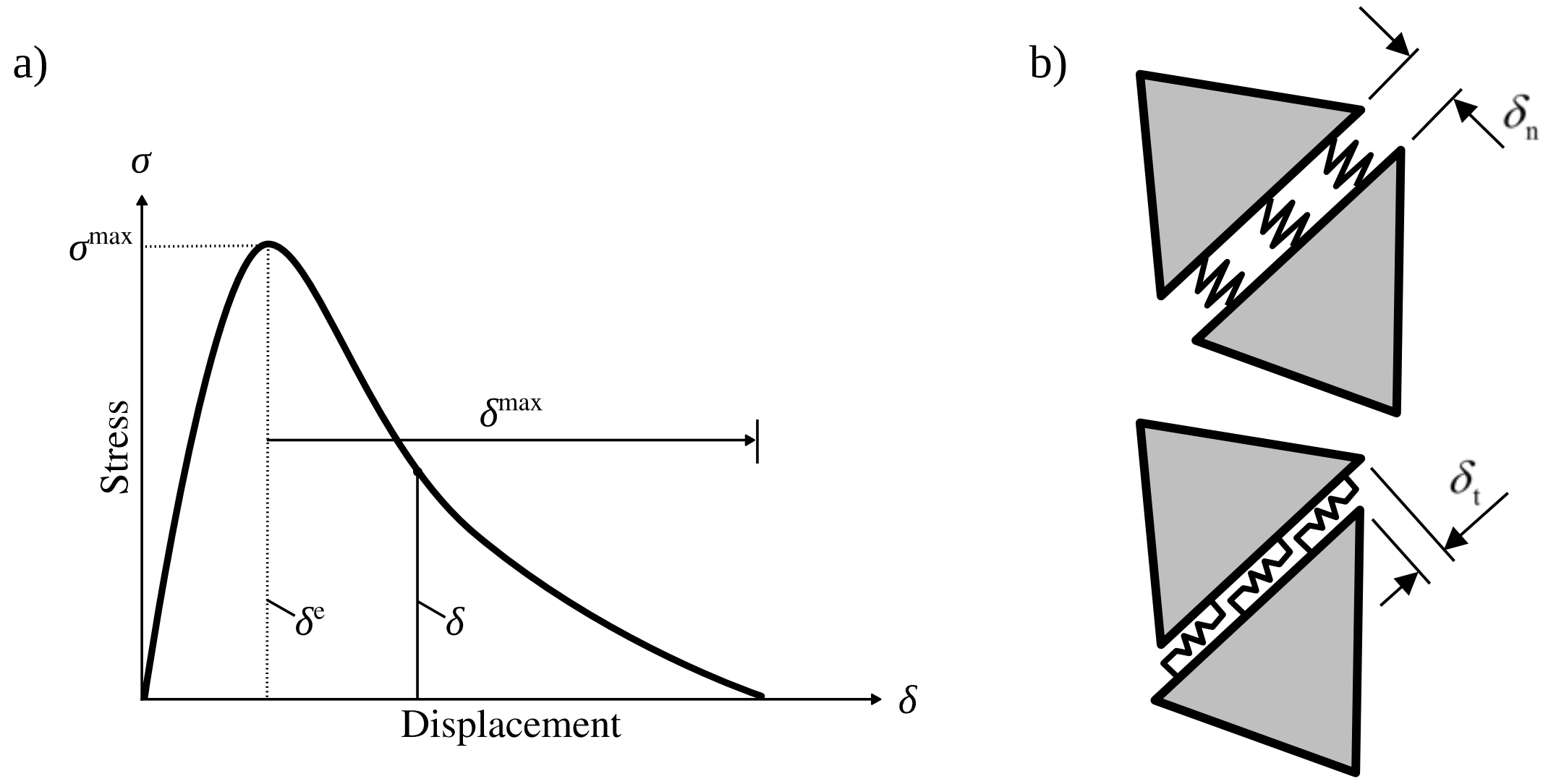

Figure 1: a) Schematic of the stress-displacement curve used to describe the behavior of an interface in FDEM. b) Schematic representation of the normal and tangential modes at the interface of two finite elements.

The damage parameter $D$ is calculated based on the amount of relative displacement that occurs after the material exits the strain-hardening regime. The relative displacement is normalized by the maximum displacement (Eq. (4)), yielding a damage parameter between 0 and 1, where 0 represents undamaged material and 1 represents a fully developed fracture.

$$D = \frac{\delta - \delta^{\mathrm{e}}}{\delta^{\max} - \delta^{\mathrm{e}}} \tag{4}$$

The material damage can be described as tensile, shear, or a mix between the two (Figure 2). The damage type is calculated by normalizing the shear and tensile displacements by their maximum allowable shear and tensile displacements, respectively. This results in a damage type ranging from 0 to 1, where 0 represents pure shear damage and 1 represents pure tensile damage.

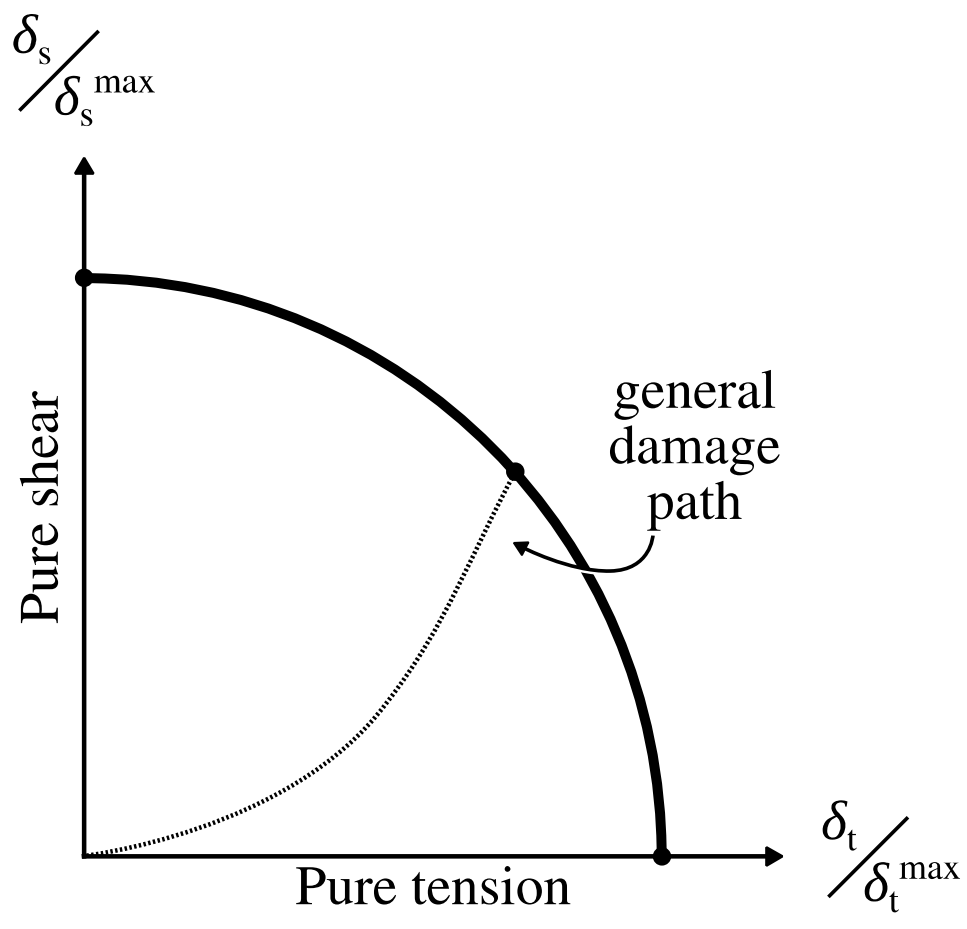

Figure 2: Illustration of the connection between tensile and shear damage for a given interface.

## 3. FDEM SIMULATIONS

**Simulation Setup**. The experimental setup used by Lee and Jeon [23] (Figure 3), is used as a reference for a 2-D numerical analysis of fracture coalescence in granite samples. Two cases were studied: a sample with a single crack oriented at different angles with respect to the loading direction (Figure 3a); and a sample with two cracks, where one crack orientation is varied with respect to the loading direction (Figure 3b). For both cases, the vertical loading is applied at a constant strain rate such that quasi-static behavior is maintained. Samples are compressed between plates, where one is given a constant velocity and the other is fixed (i.e., no translation, rotation, or rigid body motion) in space. The coefficient of friction between the sample and the plates $\mu_s$ is assumed to be 0.5.

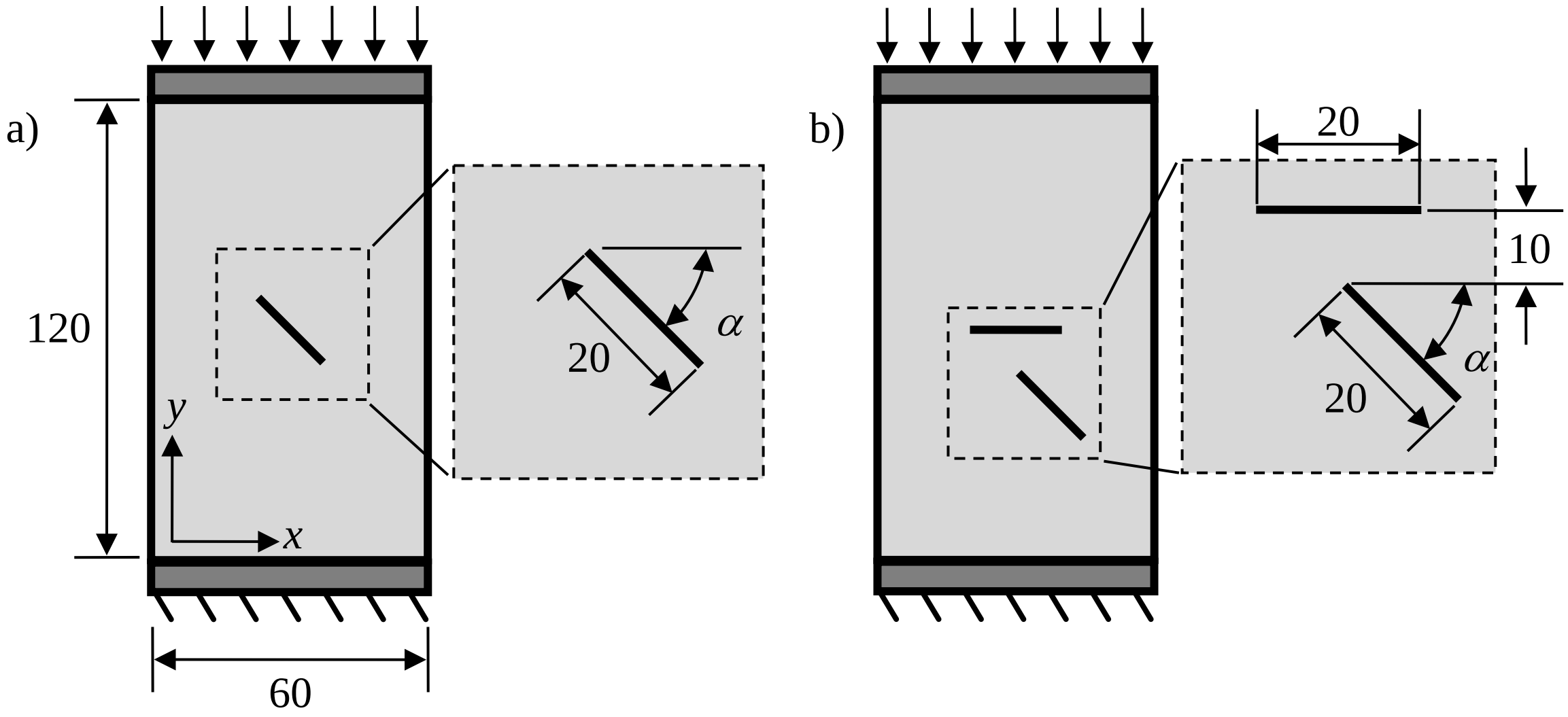

Figure 3. a) General model setup for the single-crack simulations; b) General model setup for the double-crack simulations; all dimensions are in millimeters.

## 3.1. Material Strength Model Verification

Following a process similar to the one proposed by Tatone and Grasselli [37], 2-D Brazilian Disk (BD) and Uniaxial Compressive Strength (UCS) simulations were performed in order to ensure the simulations accurately reproduce experimentally measured material behavior. The material properties used in the simulations (Table 1) were based on experimental measurements published by Lee and Jeon [23]. The UCS specimen is 120 mm x 60 mm in height and width, respectively; the BD specimen has a radius of 25.4 mm. The simulation material is considered to be homogeneous, isotropic, and under plane stress conditions.

*Table 1. Hwangdeung granite material properties reported by Lee and Jeon* [23].

| *Property* | *Value* |
|---|---|
| Young's modulus / GPa | 55 |
| Poisson's ratio | 0.15 |
| Density / (kg/$m^3$) | 2650 |
| Tensile strength / MPa | 9.2 |
| Cohesion ($c$) / MPa | 55.4 |
| Internal angle of friction ($\phi_c$) / degrees | 35 |

The UCS results, using the material properties in Table 1, adequately replicate the compressive strength reported by Lee and Jeon; the experimental and simulated UCS values are 209 and 212.1 MPa, respectively (Figure 4). Similarly, the BD simulation produces a tensile strength of 9.9 MPa (Figure 5), which closely matches the reported experimental value of 9.2 MPa. Both the BD and UCS simulations used a loading rate of 0.1 m/s and a nominal element size of 1.0 mm.

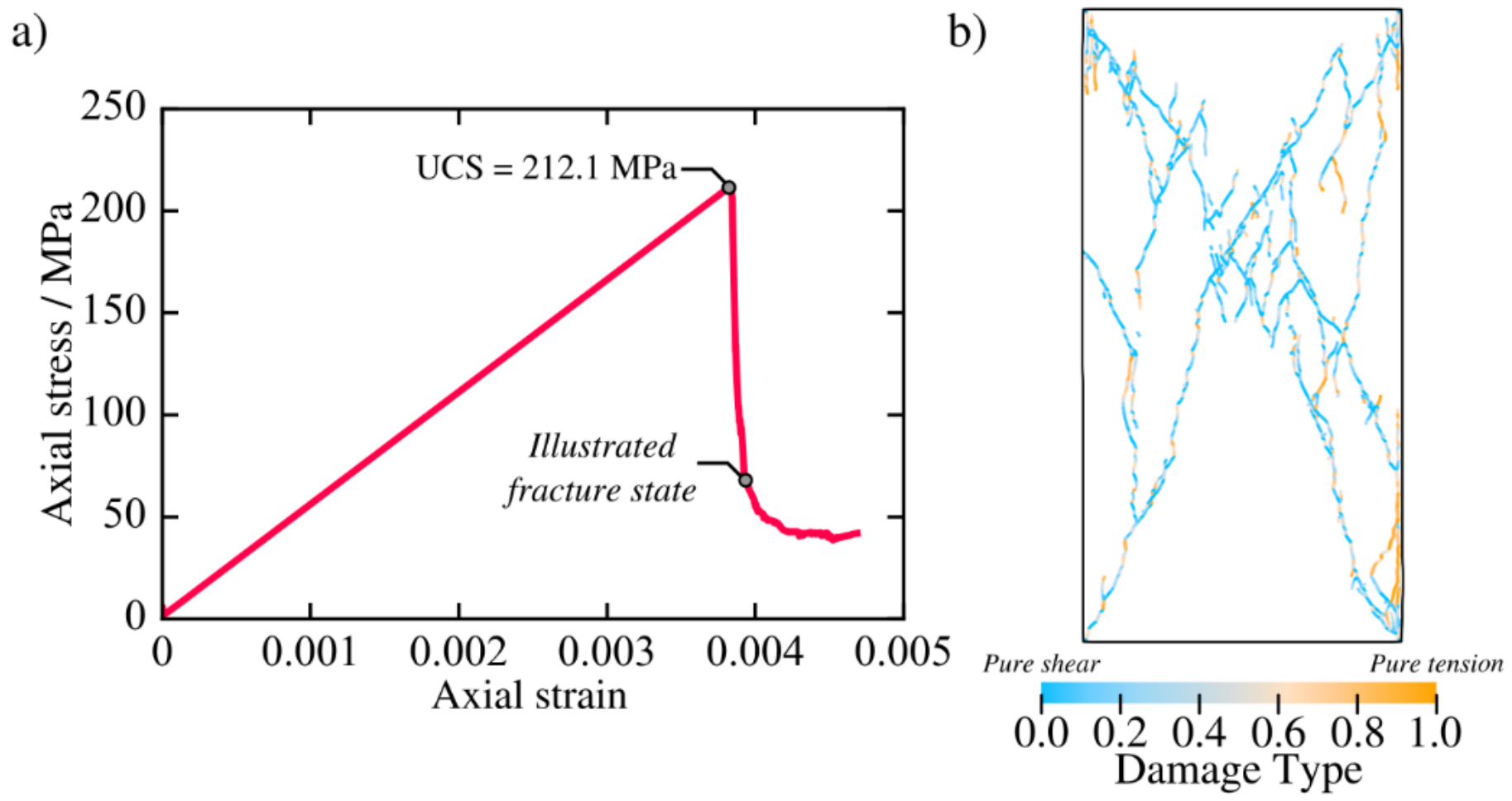

Figure 4: (a) Results of the UCS simulation including (b) fracture and damage type at the approximate moment when the sample fails.

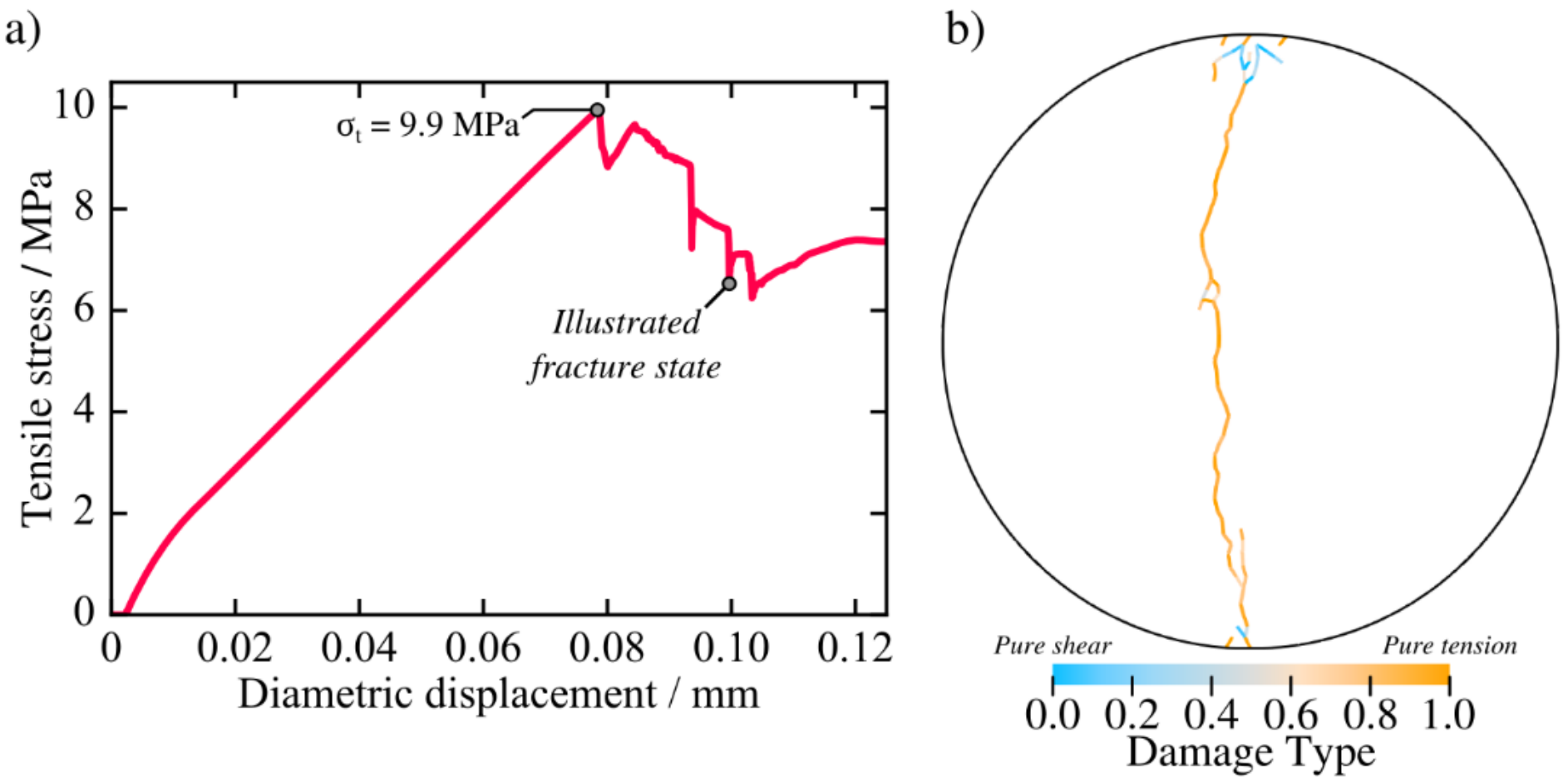

Figure 5: (a) Results of the BD simulation including (b) fracture damage and type of fracture at the approximate moment when the sample fails.

## 3.2. Results and Discussions

**Single-crack sample.** Numerical results are compared to experimentally observed fractures patterns in Figures 7-12 for the case of a single-flaw. The numerical results depict fractures color coded according to their damage type, where blue and orange are associated with shear and tensile damage, respectively (cf., Fig. 1b and 2). In most cases, only fully developed fractures (i.e., a damage value equal to 1) are visualized, as the numerical results are compared to unaided visual observations of experimental fracture patterns extracted from published images [23]. A low fracture threshold (i.e., a damage value equal to 0.1) is used when illustrating the time series progression of fracture, as small amounts of damage noticeably affect the stress field. The state of fracture illustrated in Figures 7-12 is shortly after the maximum axial stress occurs, the inclination angle of the single-flaw was varied from 0-90° in increments of 15° in these figures.

Inspection of crack propagation through time, as predicted by the HOSS numerical results (Figure 6), aids in understanding fracture coalescence mechanisms with greater fidelity than the laboratory experiments permitted. For a specimen with a pre-existing single crack inclined at 45°, the fracturing sequence starts with tensile wing cracks oriented at $2\alpha$ from the crack plane, which agrees with the Griffith elliptical crack model [38]. The fractures continue to propagate in tension with a stable manner as the uniaxial load continues to increase. As time progresses, the crack propagation begins to include more shear fracturing (blue) as newly formed tensile fractures coalesce and extend. At complete failure of the specimen, these mixed shear and tension fractures propagate towards the outer lateral boundaries of the specimen resembling a conventional UCS specimen crack geometry. Clamping stresses between end platens and the specimen with the 0.5 coefficient of friction contribute to this transition from tensile to mixed tensile and shear modes of fracture propagation. This time series prediction allows direct categorization of fractures formed without significant influence from the specimen boundary from those that formed due to interaction with the boundary. In addition, the stress state accompanying fracture formation shows that large compressive maximum principal stresses develop at the tips of the pre-existing crack. Fracturing tends to develop in regions where the maximum tensile principal stress is largest.

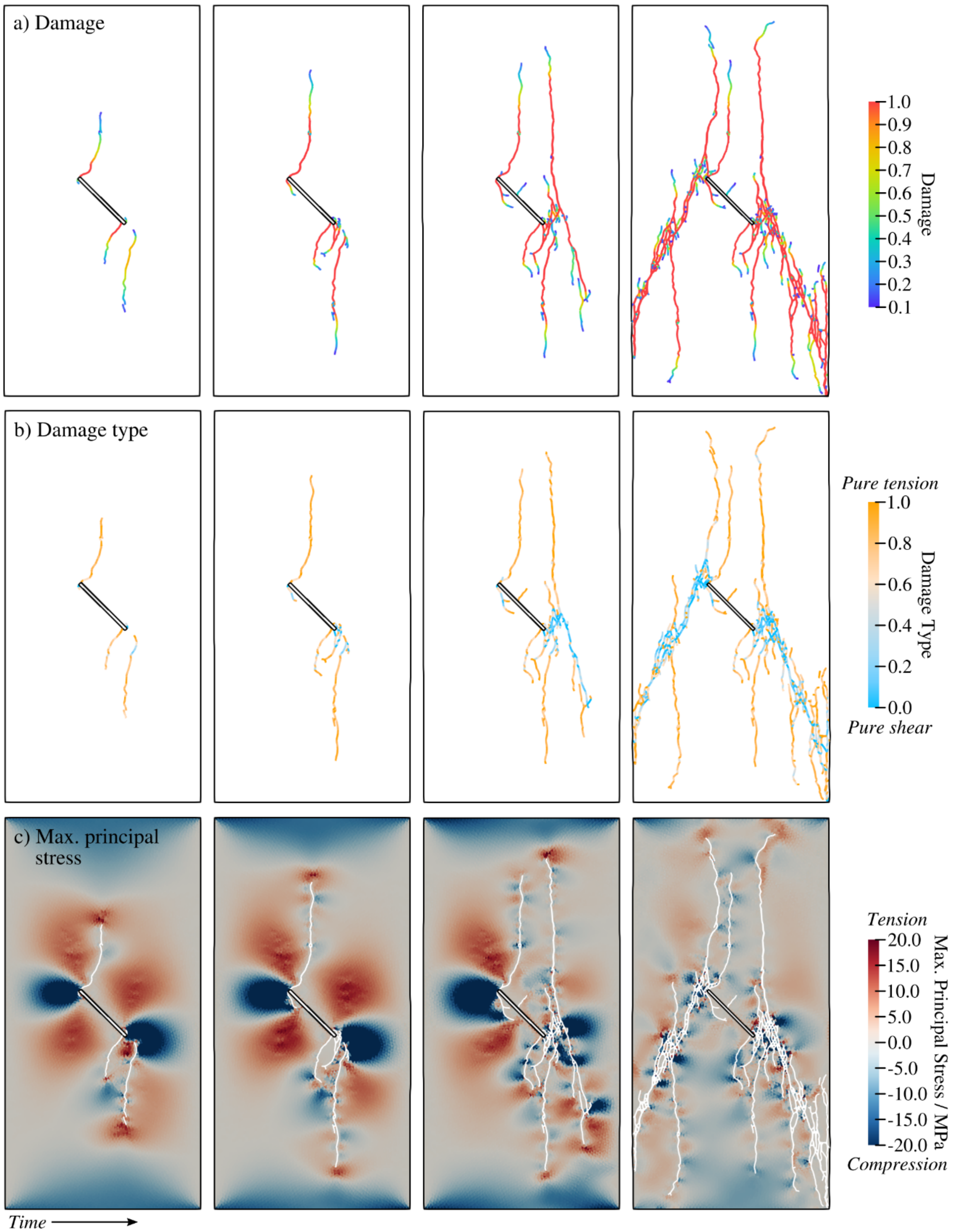

Figure 6: (a) Damage progression, (b) damage type, and (c) maximum principal stress in a sample with a pre-existing crack at an inclination angle $\alpha = 45°$.

When the flaw orientation is $\alpha = 0°$ (Figure 7), tensile cracks emanate from the center of the flaw and travel in the direction of the major principal stress. Similarly, after the onset of damage for the 15-90° flaw orientations, tensile fractures propagate toward the direction of major principal stress. As the inclination angle of the flaw increases, the specimens experience greater amounts of shear fracturing until the specimen eventually fails in shear (Figure 11).

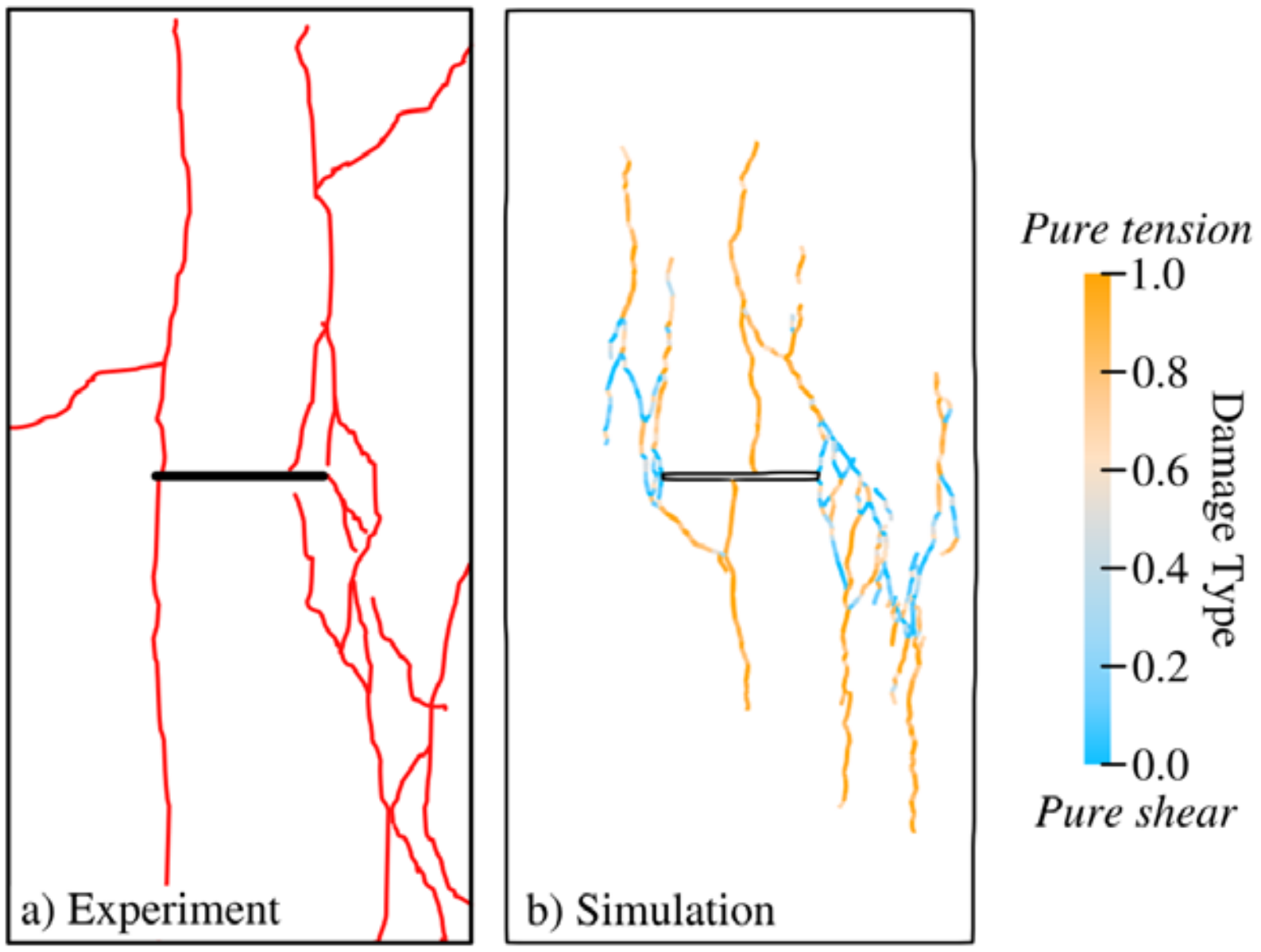

Figure 7: Single-flaw specimen with a flaw inclination angle of $\alpha = 0°$.

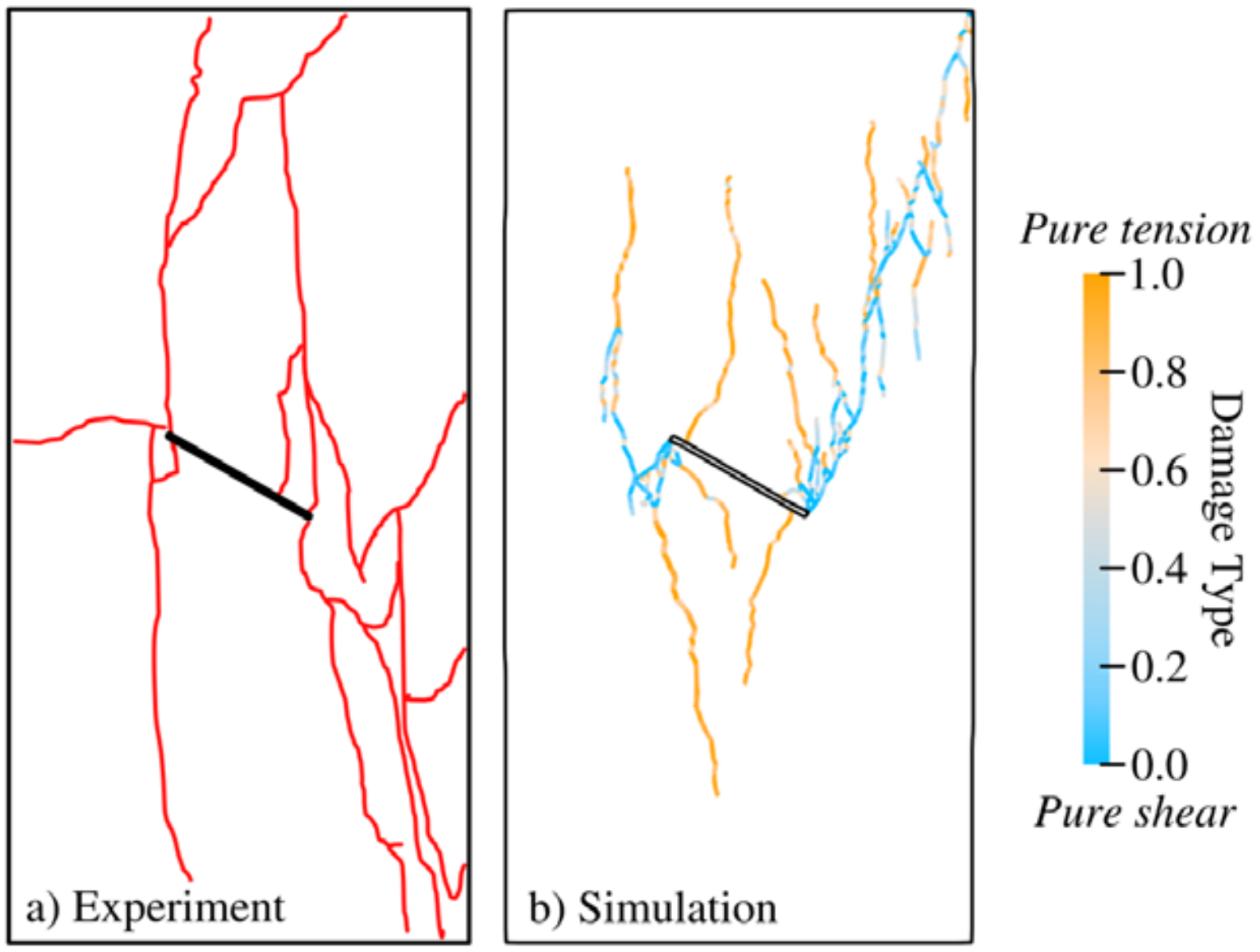

Figure 8: Single-flaw specimen with a flaw inclination angle of $\alpha = 30°$.

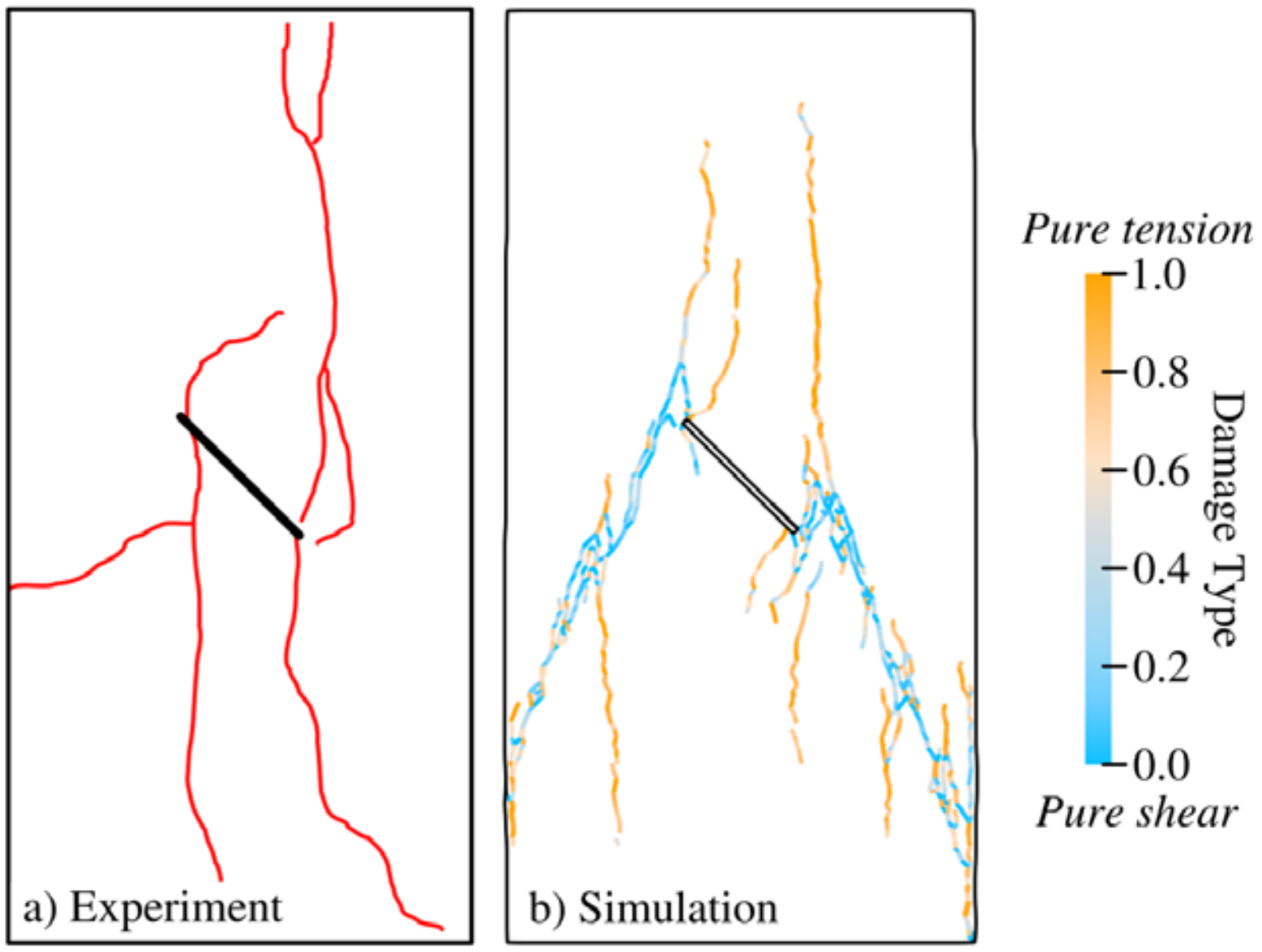

Figure 9: Single-flaw specimen with a flaw inclination angle of $\alpha = 45°$ .

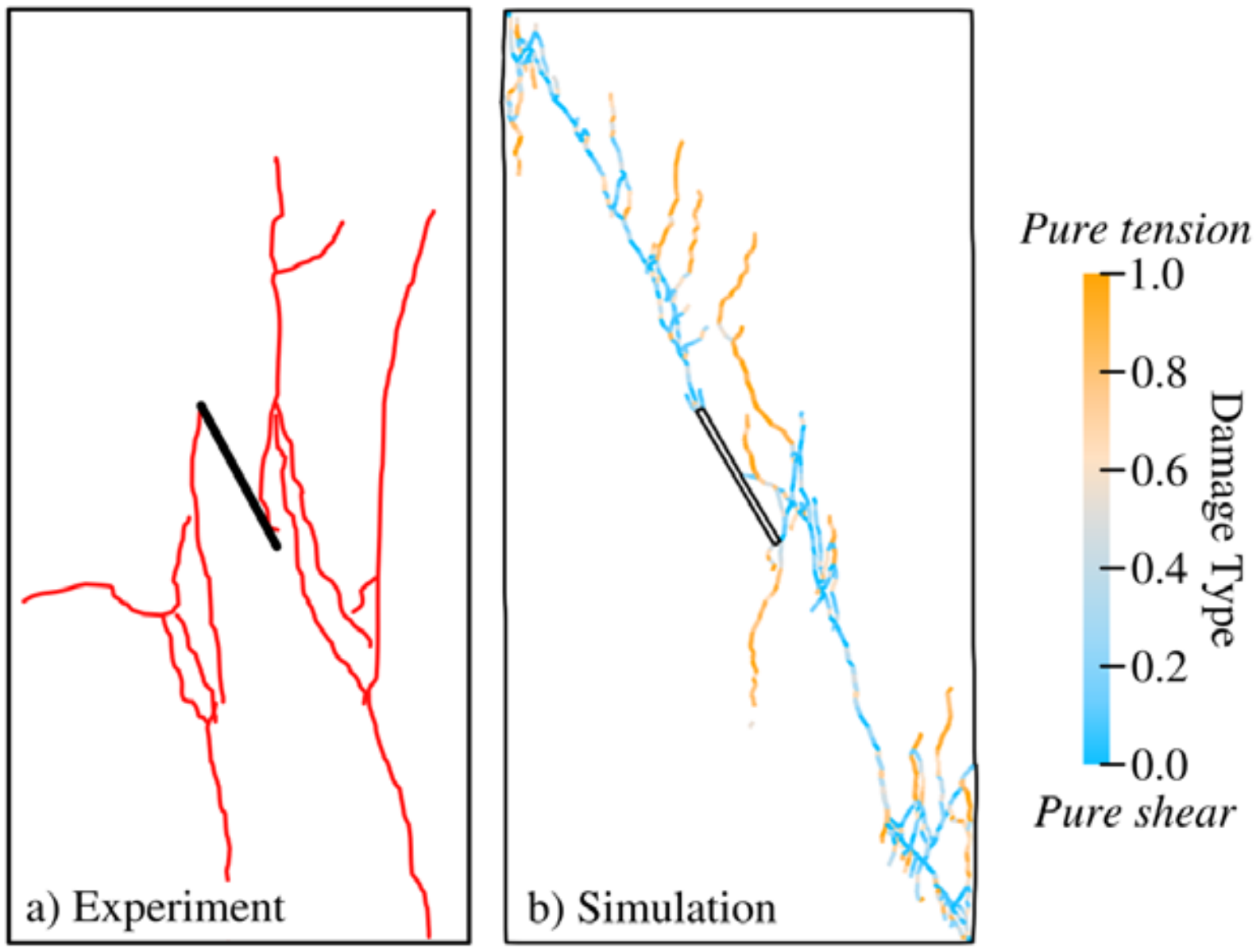

Figure 10: Single-flaw specimen with a flaw inclination angle of $\alpha = 60°$ .

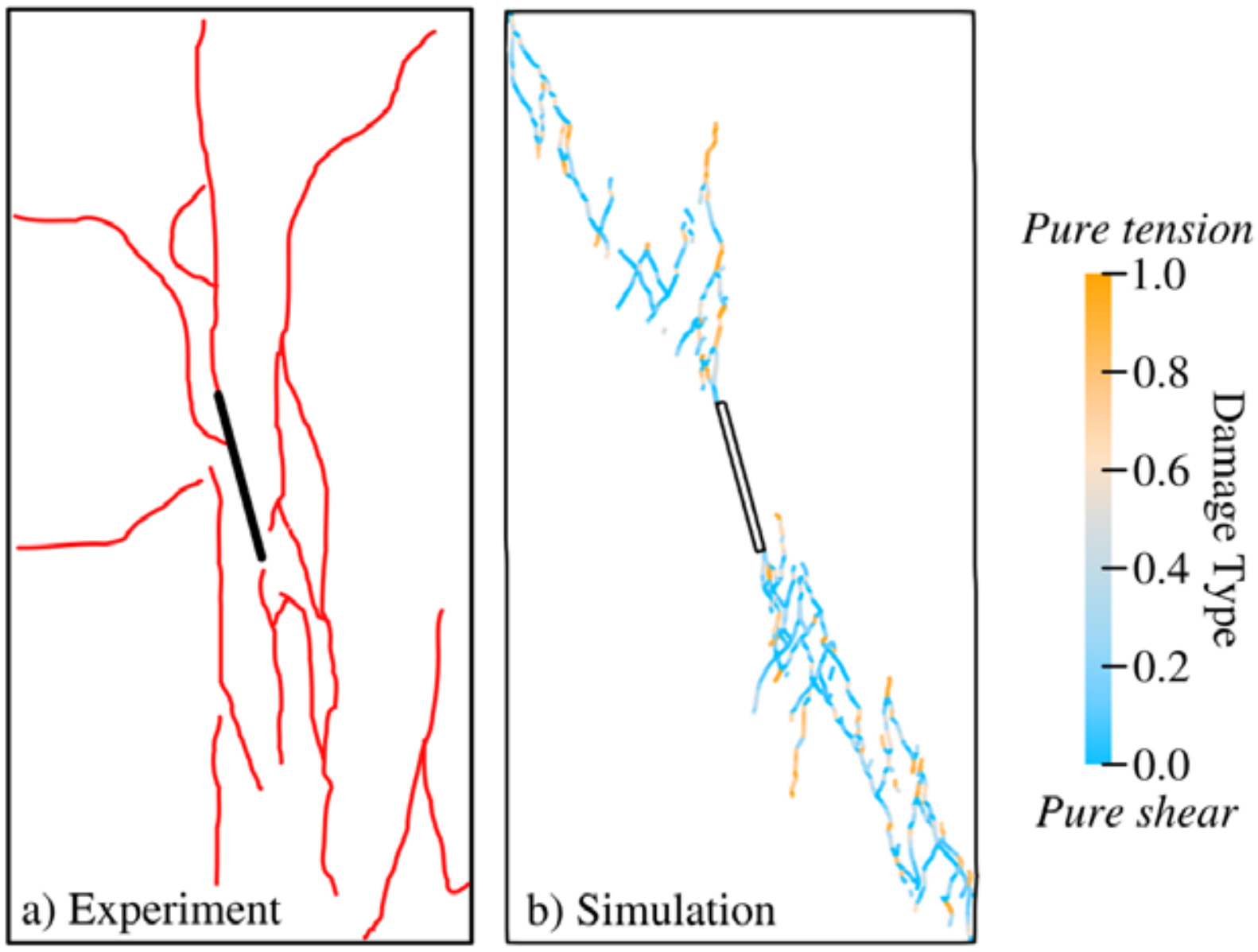

Figure 11: Single-flaw specimen with a flaw inclination angle of $\alpha = 75°$.

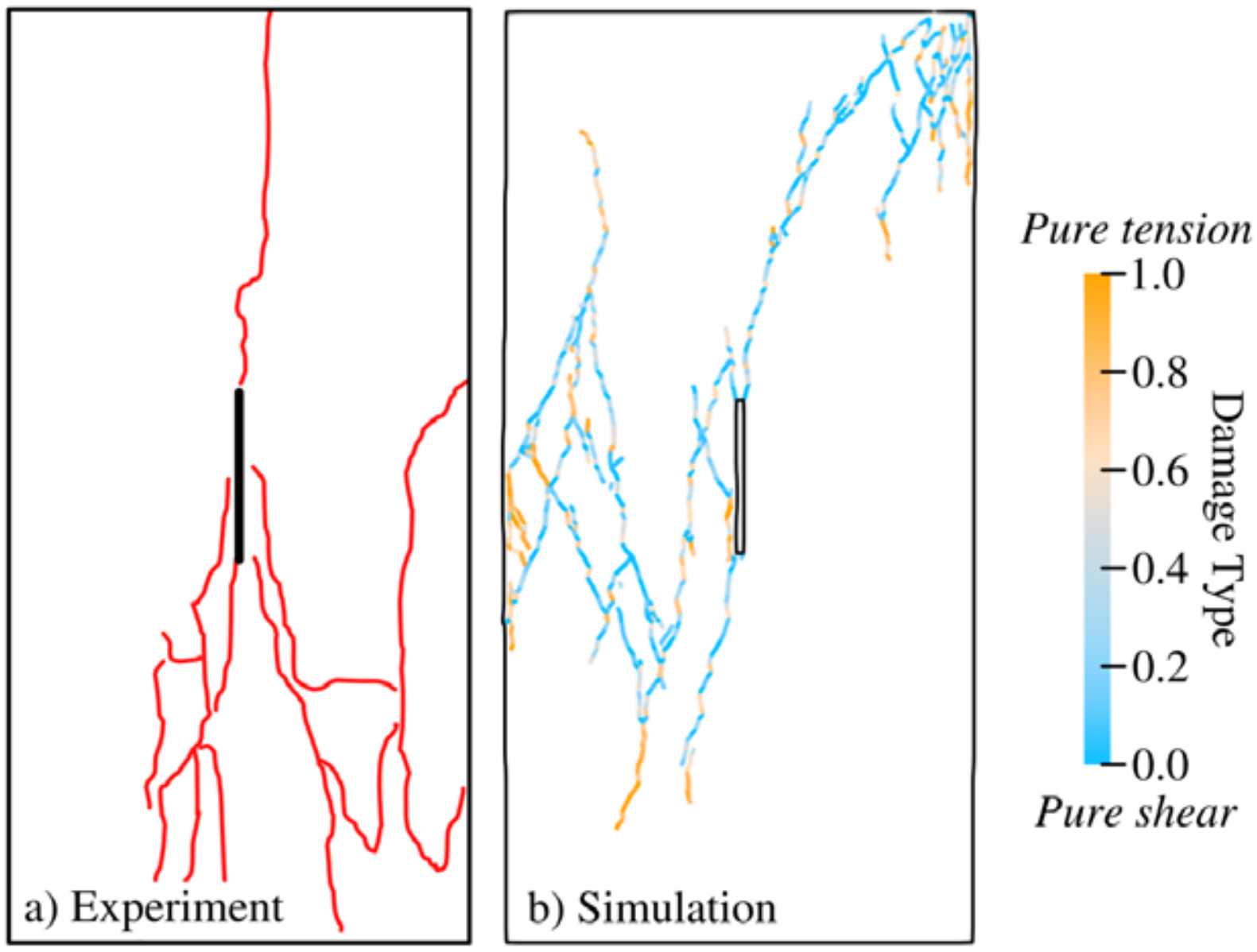

Figure 12: Single-flaw specimen with a flaw inclination angle of $\alpha = 90°$.

The average stress in the sample (Figure 13) was calculated based on the force applied to the sample by the plates divided by the cross-sectional area of the sample. The stress profile in the linear-elastic regime is not identical for all samples due to the variation of fracture initiation for each specimen. Figure 14 shows the peak stress as a function of the inclination angle $\alpha$. As the angle of inclination increases, the peak stress also increases. The simulations produce peak stress values that are consistent with the experimental observations of peak stress and flaw orientation. While the DEM simulations performed by Lee and Jeon [23] produced a similar trend, the current model more accurately predicts the magnitude of the stress, as represented in Figure 14.

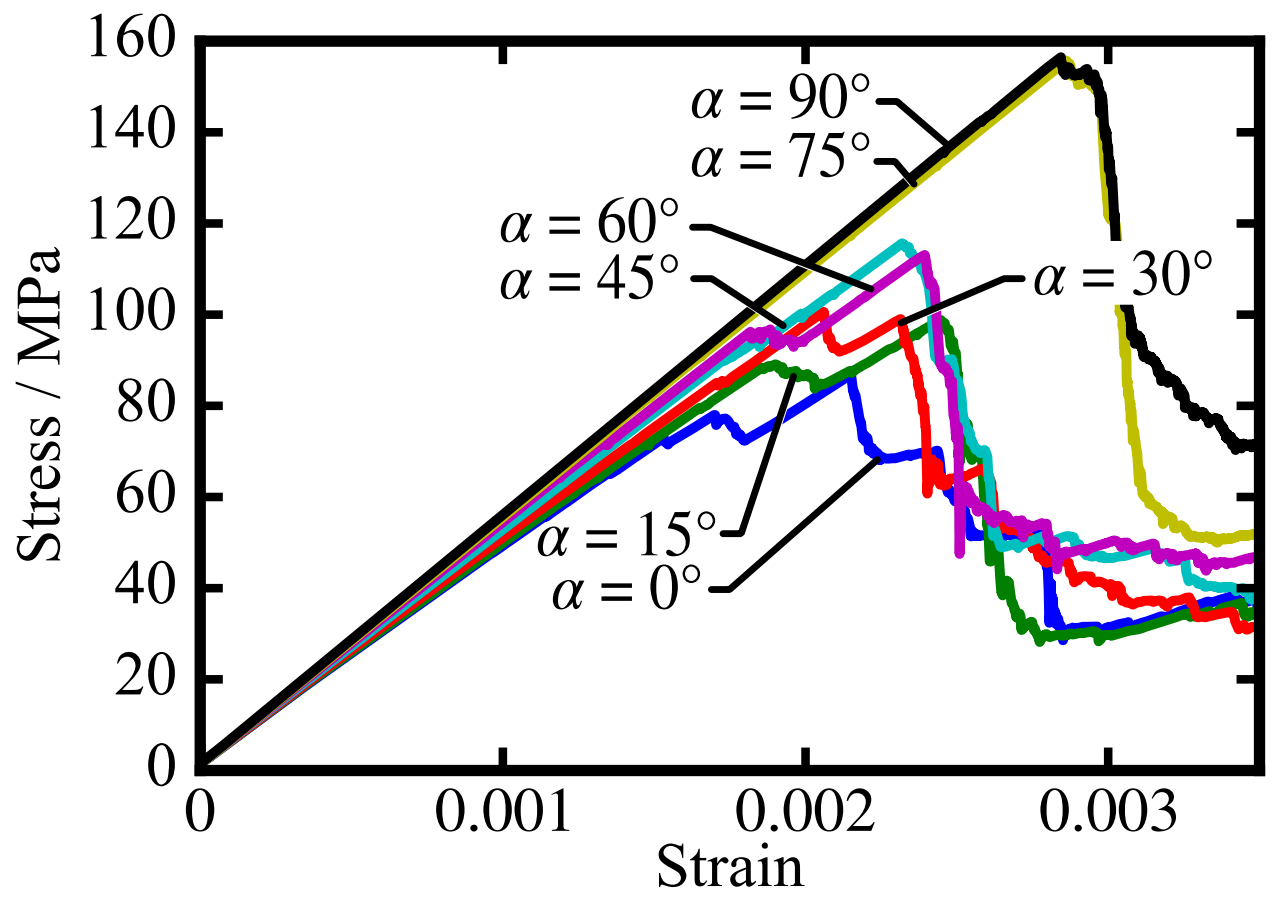

Figure 13: Axial stress-strain behavior of the single-flaw virtual specimens

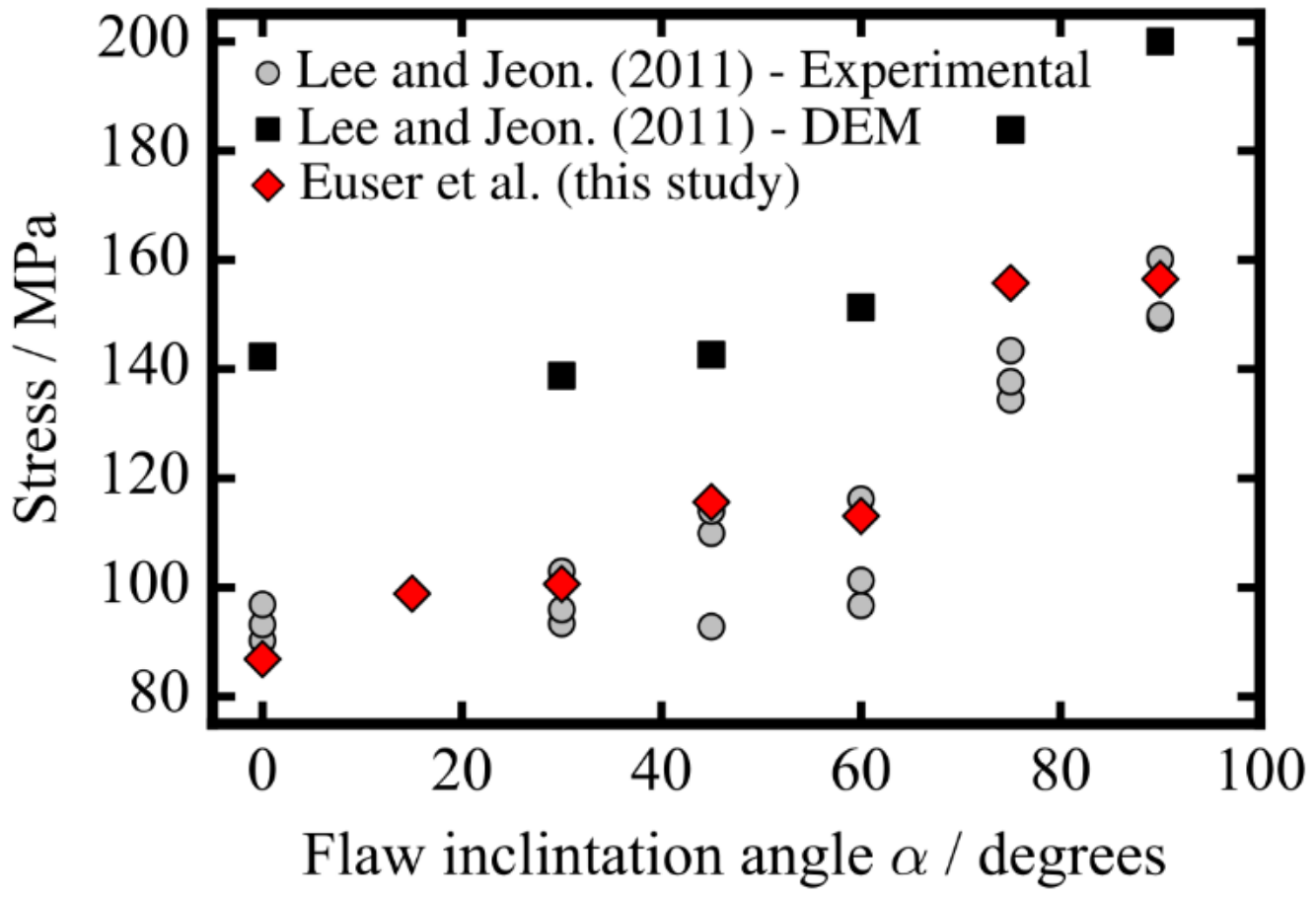

Figure 14: Peak axial stress as a function of flaw inclination angle for the single-flaw specimens. The numerical results are compared to experimental and numerical data reported by Lee and Jeon [23].

**Double-crack sample.** Figures 16 – 20 illustrate comparisons of the numerical and experimental results for specimens with two flaws. The inclination angle of the second flaw (Figure 3d) is varied from 15-90° in increments of 15°. Figure 15 illustrates the temporal evolution of the fracture in a double-crack specimen. Initially, tensile cracks appear near the center of the horizontal crack and propagate in the direction of the maximum principle stress, and wing cracks emanate from the end of the inclined crack. As time progresses, shear fractures nucleate on the left edge of the horizontal flaw and near the right tip of the inclined fracture. The shear fractures that nucleated near the right tip of the inclined flaw propagate parallel to the flaw orientation, following a path through a region of large tensile principal stress.

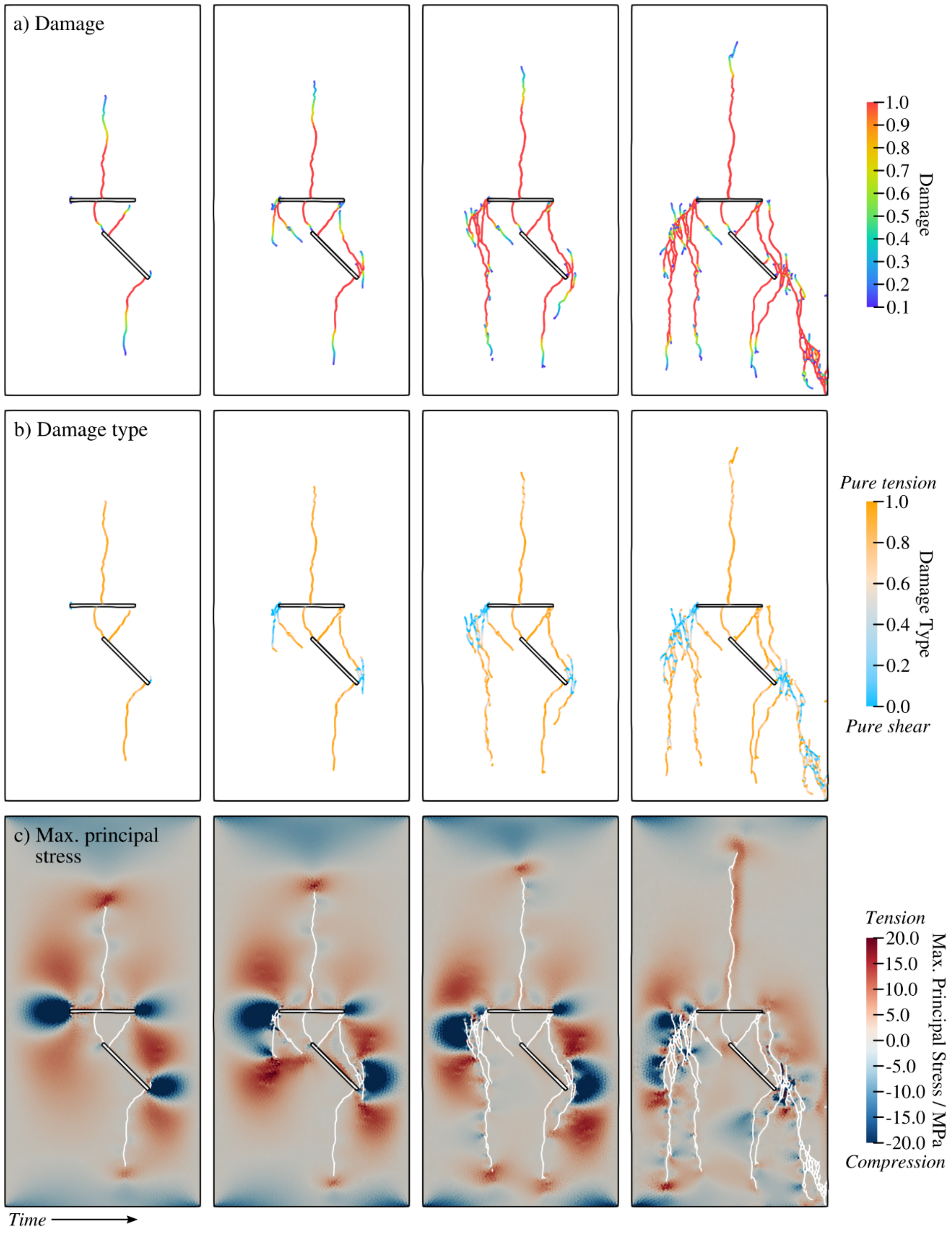

Figure 15: (a) Damage progression, (b) damage type, and (c) maximum principal stress development in a double-crack specimen with an inclination angle of the secondary crack $\alpha = 45°$.

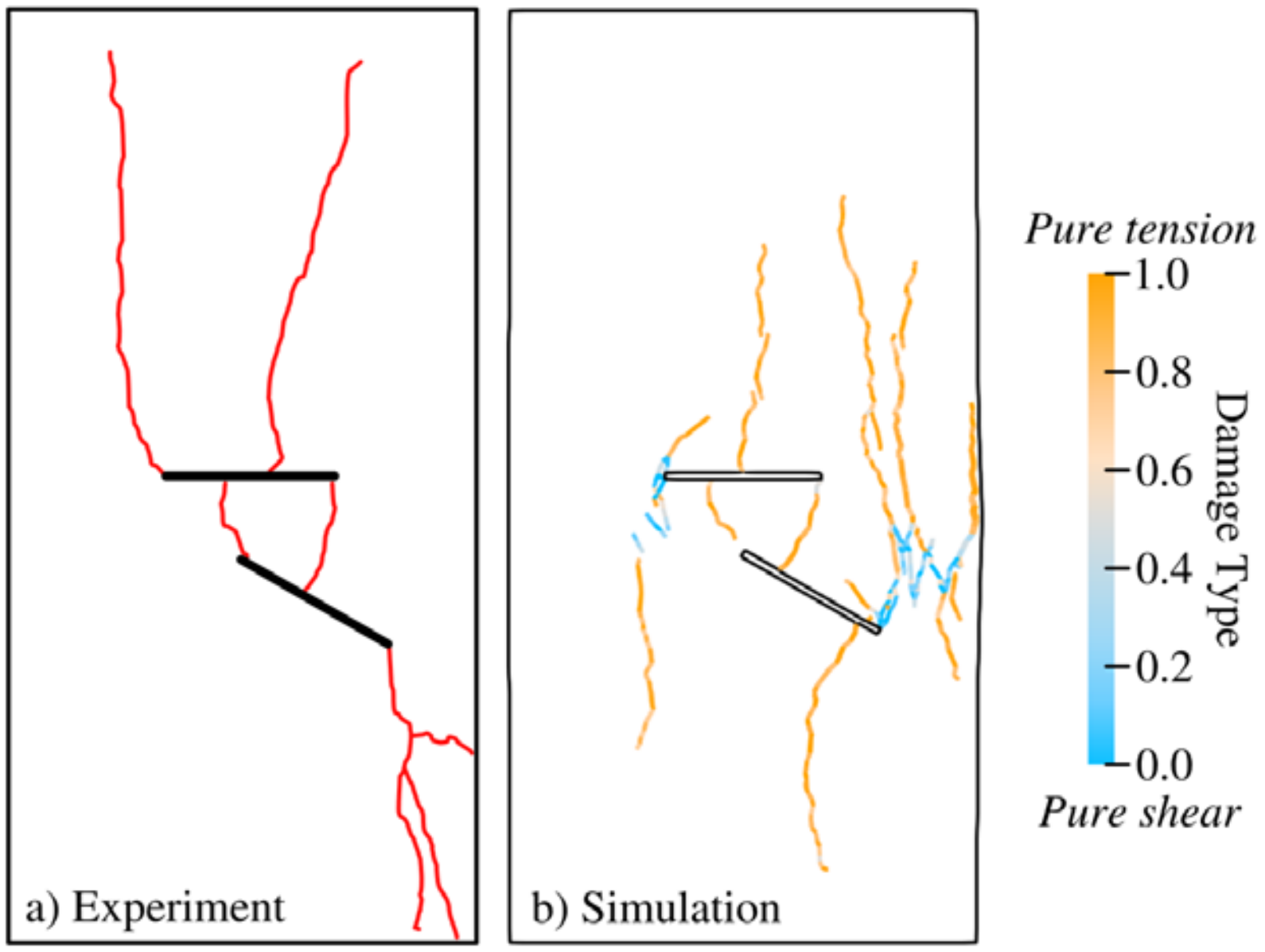

Figure 16: Double-flaw specimen with a flaw inclination angle of $\alpha = 30°$.

For all cases (Figures 16 – 20), mode I opening near the center of the horizontal crack is always observed. In addition, nearly all simulations experience a small amount of shear fracture on the left tip of the horizontal crack, resulting in tensile fractures that propagate towards the bottom of the sample. When $\alpha = 30°$ (Figure 16), prominent features of the experimental specimen can be observed in the numerical results, particularly the tensile cracks that connect the two pre-existing cracks. The mixed-mode fracture emanating from the right tip of the inclined flaw is qualitatively similar to the experimental fracture patterns. However, the simulation does not reproduce the predicted vertical fracture extension emanating from the left tip of the horizontal flaw.

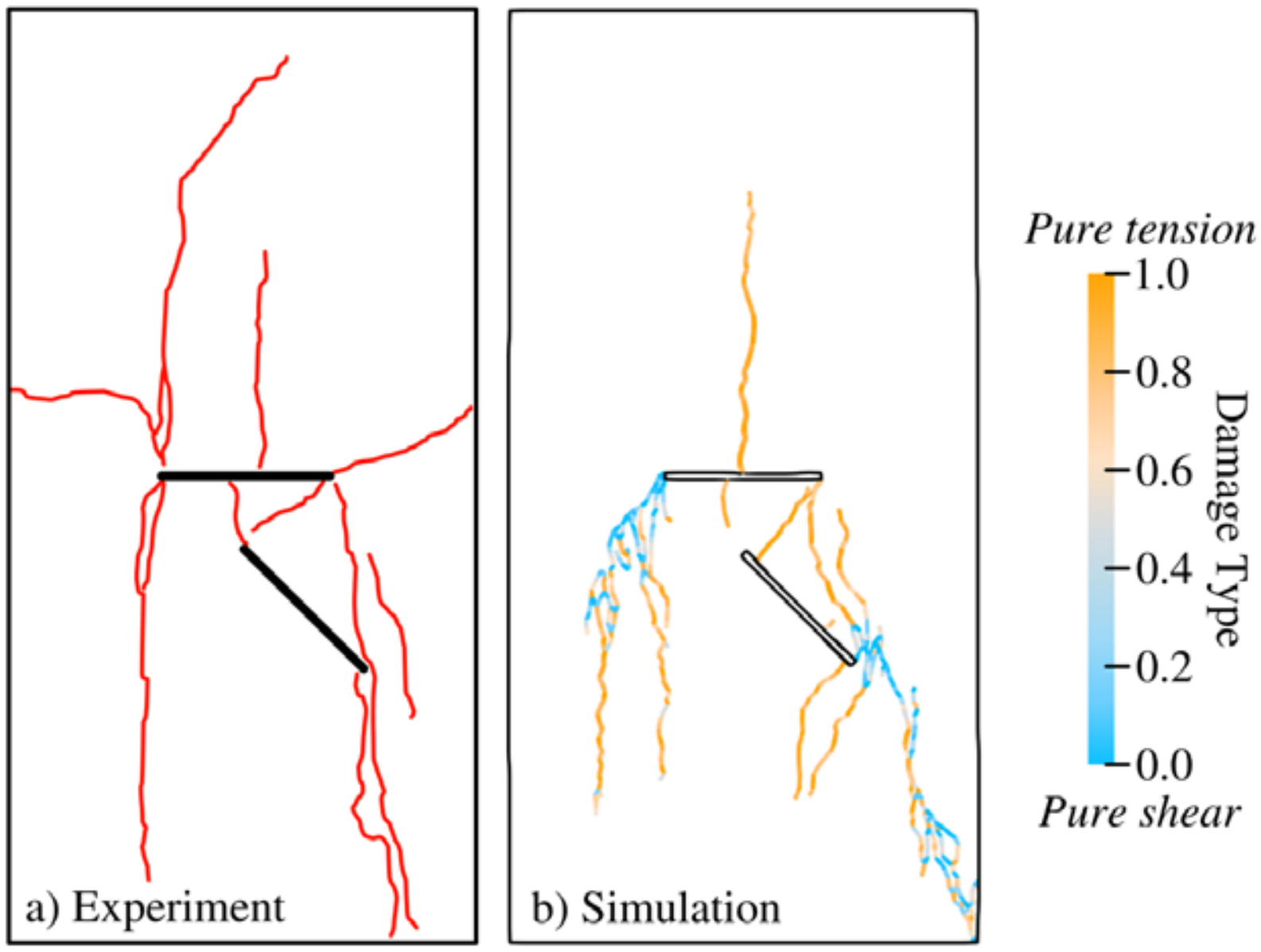

Figure 17: Single-flaw specimen with a flaw inclination angle of $\alpha = 45°$ .

When the inclined crack is oriented at an angle of $\alpha = 45°$ , a connection is formed between the right tip of the horizontal flaw and both tips of the inclined flaw. The numerical model is similar to physical observations in many aspects, although the failure plane of the virtual specimen is somewhat different from the real specimens; the simulations tend to predict fractures intercepting the sides of the specimen rather than the ends of the specimen, as found in many of the experiments.

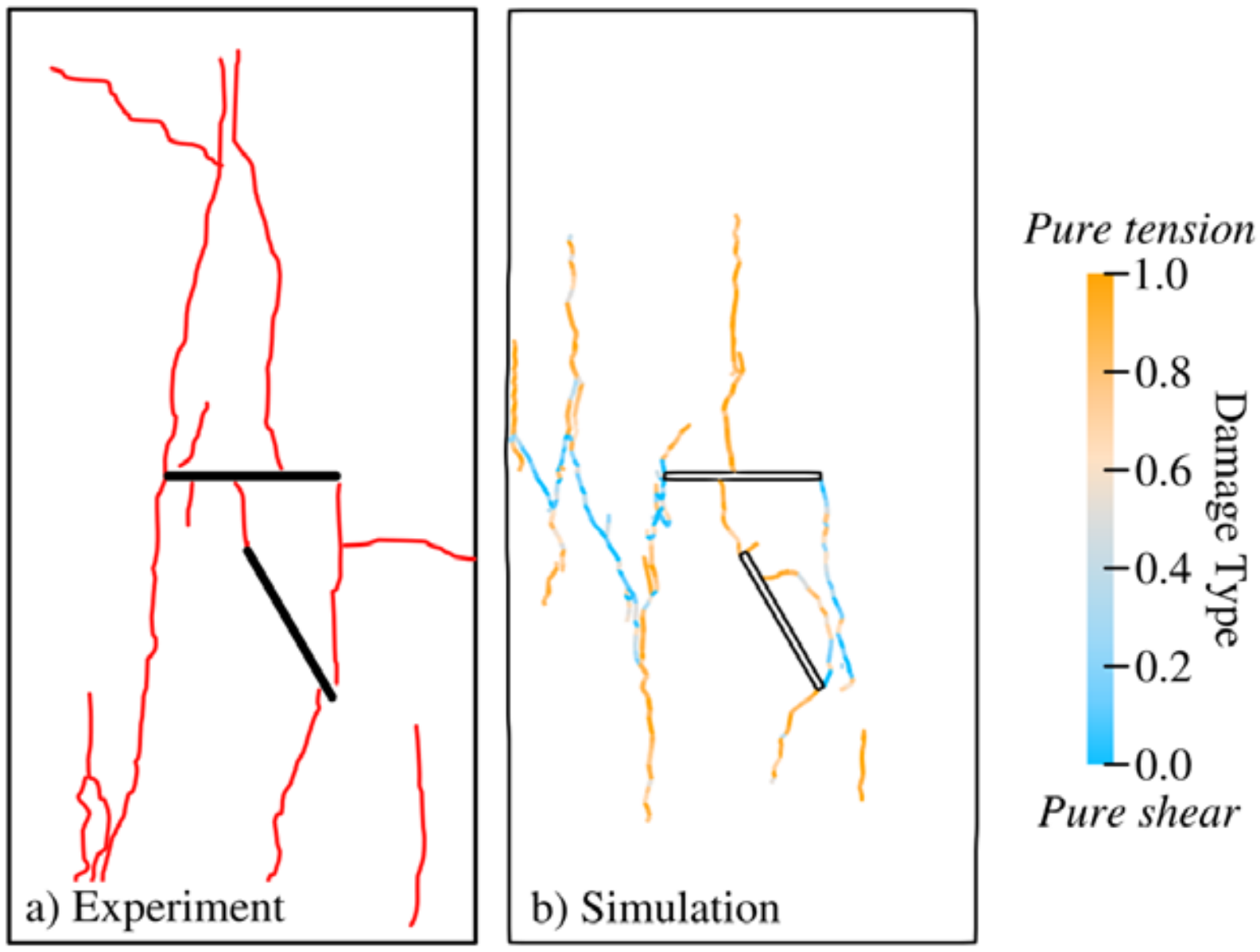

Figure 18: Single-flaw specimen with a flaw inclination angle of $\alpha = 60°$ .

For an inclination angle $\alpha = 60°$ (Figure 18), the tensile fracture originating from the center of the horizontal flaw connects to the left tip of the inclined flaw, while mixed-mode fractures propagate from the right tip of horizontal flaw to the right tip of the inclined flaw in a similar manner to the experiments. However, the simulation again predicts development of shear fractures to the sides of the specimen while this is not observed in the experiments by simple visual inspection of the sample.

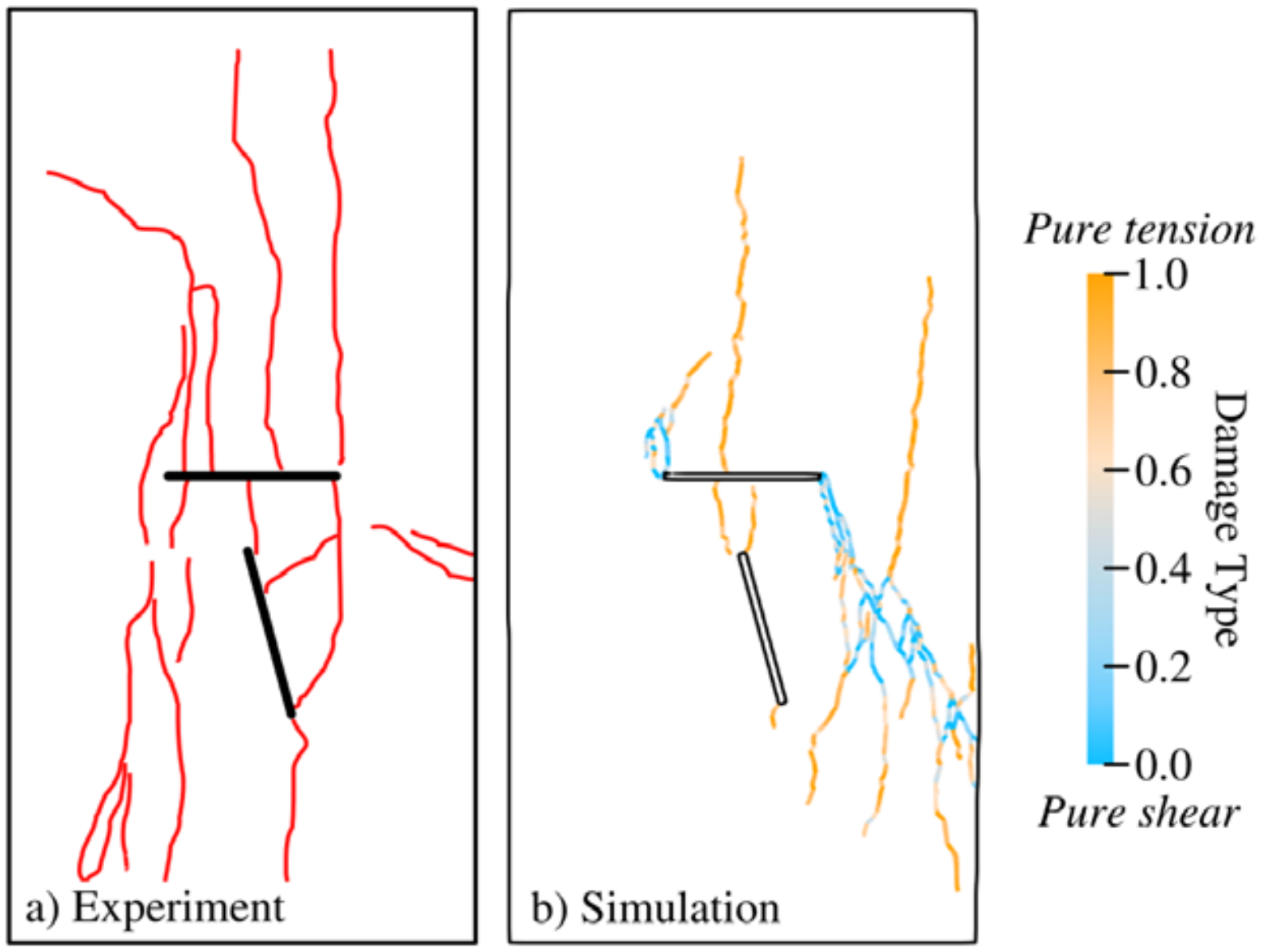

Figure 19: Single-flaw specimen with a flaw inclination angle of $\alpha = 75°$.

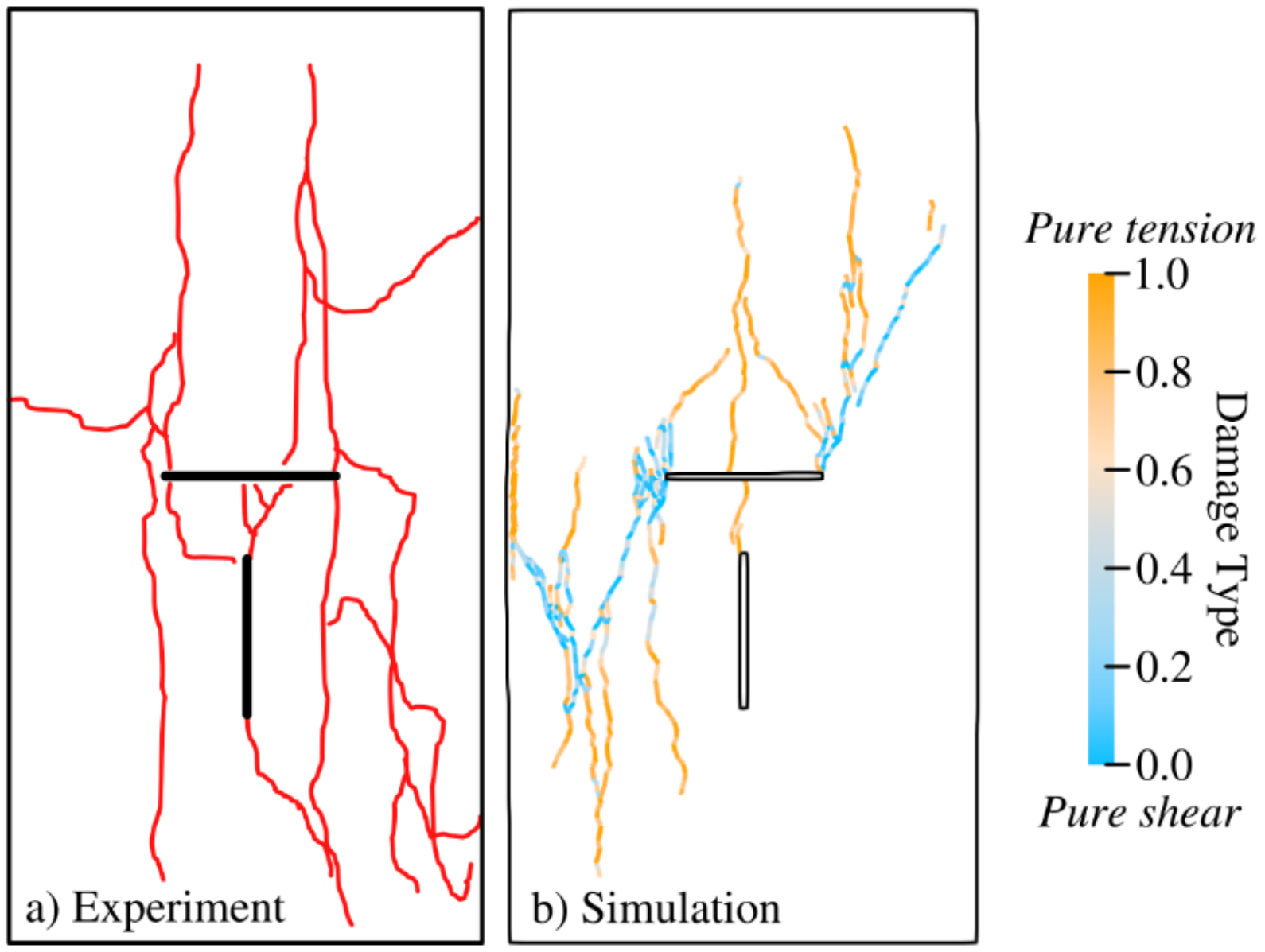

Figure 20: Single-flaw specimen with a flaw inclination angle of $\alpha = 90°$.

The fracture patterns for inclination angles of $\alpha = 75°$ and 90° (Figures 19 and 20) both feature tensile cracks that develop between the center of the horizontal flaw and the neighboring tip of the inclined flaw. Note that the tensile dominated fractures (orange) more closely match the experimental fracture patterns and that tensile fractures will be more visible than shear fractures due to larger aperture, so the locations of the shear fractures in the experiments could have existed to some degree, but were undetected by the naked eye.

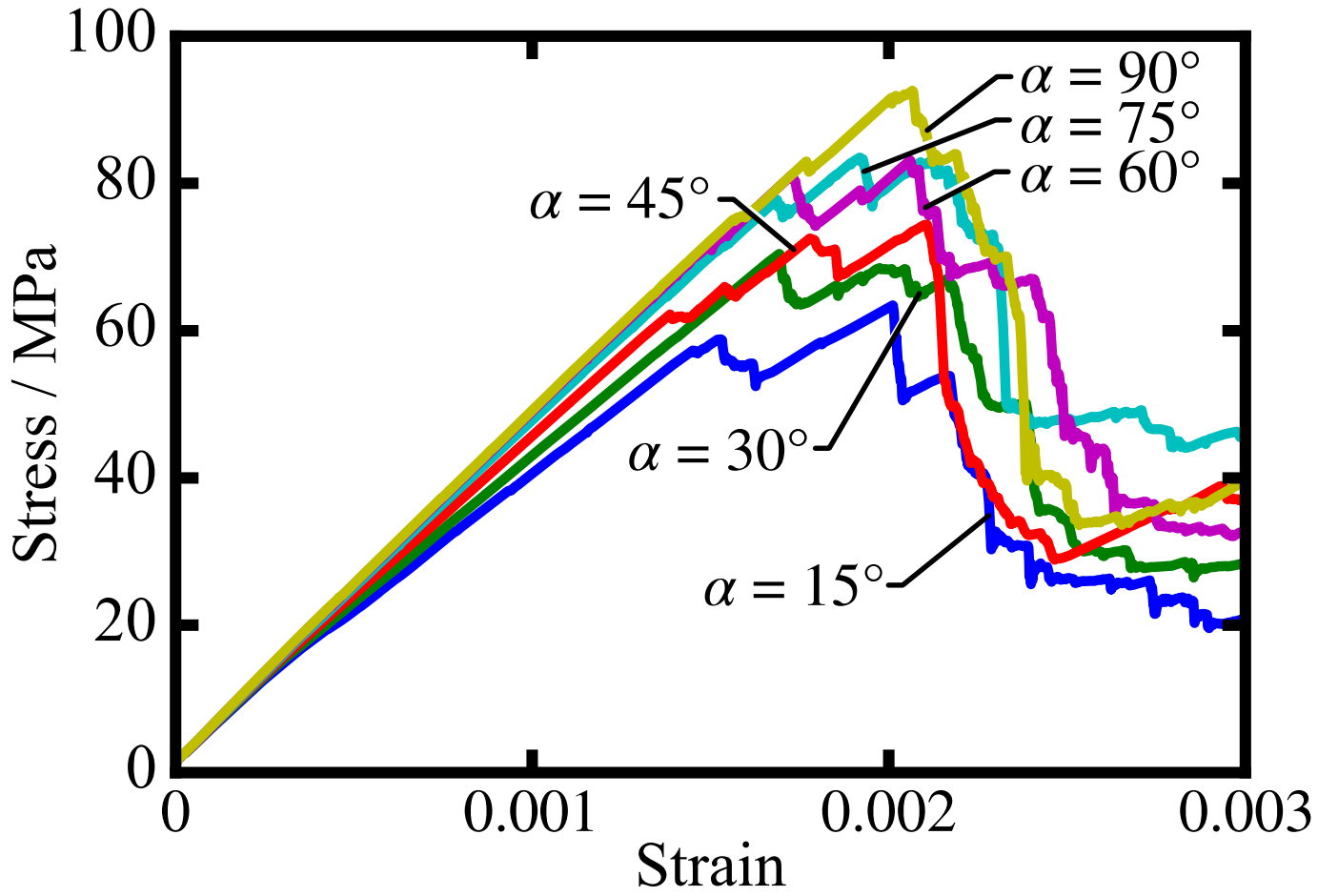

Figure 21: Axial stress-strain behavior of the double-flaw virtual experiment.

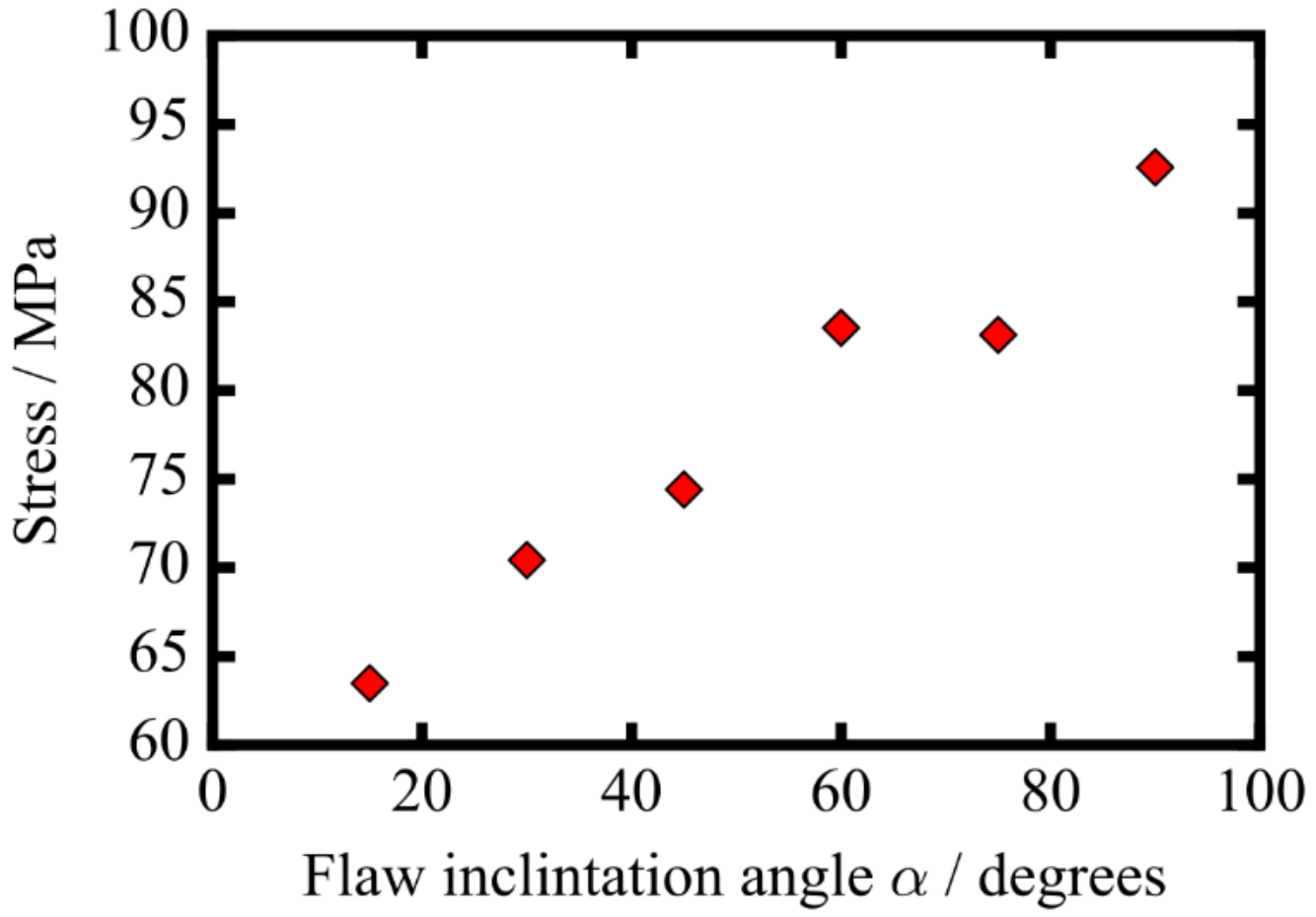

Figure 22: Peak axial stress as a function of flaw inclination angle for the double-flaw virtual specimens.

Figure 21 illustrates the evolution of stress in the double-crack samples as a function of strain for different flaw orientations. Each specimen experiences catastrophic failure at roughly 0.2% axial strain, which coincides with the axial strain necessary for catastrophic failure in a specimen with a single crack oriented perpendicular to the loading direction ($\alpha = 0°$). Figure 22 shows the peak stress as a function of the inclination angle $\alpha$ for the double-flaw specimen; however, no experimental data are reported by Lee and Jeon [23] for comparison. Similar to the case of the single-crack, the peak stress increases with increasing inclination angle. The peak stresses are lower than those observed for the single-flaw, which can be attributed to the presence of the primary horizontal flaw. As the secondary flaw becomes more vertical (i.e., $\alpha = 60° - 90°$), the horizontal crack shields the secondary crack against the vertical loading, and the magnitude of the peak stress (90 MPa) is consistent with the peak stress of the horizontal single-crack simulation ($\alpha = 0°$).

**Discussion.** For the single-crack specimens, it is clear that there is a relationship between the flaw inclination angle and the amount of shear fracturing observed within the specimens. For lower inclination angles, a mix of tensile and shear fractures contribute to the failure of the specimen. As the angle of inclination $\alpha$ increases, the failure of the specimens can be predominantly associated with shear fractures that are quasi-coplanar to the inclination angle. On the other hand, the double-crack specimens largely experience failure due to shear fractures that nucleate at the tips of the primary and secondary cracks. Note that, as the angle of inclination increases for the secondary crack in the double-crack specimen, shear fracture nucleation shifts from the right tip of the secondary flaw to the right tip of the primary (i.e., horizontal) flaw. In both the single- and double-crack cases, tensile fractures are generally ($\alpha < 60°$ for the single-crack specimens) more predominant than shear fractures in the early stages of fracture propagation. As time progresses, more and more shear fractures develop in the simulation leading to the eventual failure of the samples in shear. In practice, it is challenging for experimental studies of fracture coalescence to determine how fractures develop over time and how these fractures subsequently affect the overall behavior a material. Identifying these correlations is key to understanding fracture processes. For example, Figures 13 and 21 illustrate the stress-strain behavior of the single- and double-crack specimens. From these figures it is clear that each specimen experiences changes in slope in the linear elastic regime, a behavior that is related to the differences in the onset of fracturing as the crack inclination angle increases.

Numerical methods such as FEM and DEM are capable of modeling fracture coalescence processes, but there are shortcomings associated with these methods. In order to accurately describe strain localization processes in geomaterials with the FEM, a very fine mesh is necessary; as such, the computational cost can become quite excessive. Similarly, DEM needs a large amount of computational resources, in addition to requiring material model calibration, in order to accurately describe fracture coalescence processes. FDEM is inherently capable of providing information concerning the temporal and spatial evolution of both fracture and stress, opening many possibilities into studying the relationship between fracture formation and changes in material behavior. At the same time, FDEM requires minimal calibration and reasonable levels of computational resources in order to closely approximate experimental material behavior.

## 4. CONCLUSION

The simulations presented in this paper demonstrate HOSS's ability to reproduce key experimentally observed fracture features. These simulations including unconfined compression strength, Brazilian tensile strength, peak compressive strength in specimens with a single crack oriented at varying angles, and fracture geometries produced in experiments with single and double flaws. The simulation provides predictions of strength in the double-flaw experiments that are consistent with the single-flaw experiments and show how strength is reduced by the presence of the second flaw where it is not shielded by the primary flaw. The simulations also allow insight into fracture processes that are not easily obtained in the experiments such as the temporal evolution of the fracture system as well as the distribution and timing of tensile versus shear fracture development. The simulations also serve as a continued validation and verification of HOSS's FDEM implementation [2, 3, 30]. Both peak stress and fracture patterns observed for single-crack specimens closely match experimentally measured values.

It has also been demonstrated that HOSS can provide the capability of illuminating details about the propagation sequence of fracture patterns that are observed in experiments, while subsequently illustrating the state of stress in the sample. In addition, the influence of end effects and their contribution to fracture interaction can also be studied in detail. FDEM models element deformation and discrete fracturing using directly measured rock properties, obtained with standard testing methods, and require only minimal calibration in order to closely approximate experimental material behavior. This validation study improves confidence that HOSS FDEM is capable of producing information to better understand the processes underlying crack initiation, propagation, and coalescence in geologic materials.

ACKNOWLEDGEMENTS

Support provided by the Department of Energy (DOE) Basic Energy Sciences program (DE-AC52-06NA25396) for the financial support. The authors would also like to thank the LANL Institutional Computing program for their support in generating data used in this work.